\input{aipcheck}

\documentclass[
  ,final            
  ,draft            
  ,numberedheadings 
  ]
  {aipproc}

\layoutstyle{8x11double}

\newcommand{\un}[1]{\ensuremath{\, \mathrm{#1}}}

\begin{document}

\title{Prospects for a radio air-shower detector at South Pole}


\author{Sebastian B\"oser\footnote{sboeser@physik.uni-bonn.de}~~for the ARA and
IceCube collaborations}{
  address={Physikalisches Institut, Universit\"at Bonn, 53113 Bonn}
}



\begin{abstract}
IceCube is currently not only the largest neutrino telescope but
also one of the world's most competitive instruments for studying cosmic rays in
the PeV to EeV regime where the transition from galactic to
extra-galactic sources should occur.  Further augmenting this observatory with
an array of radio sensors in the 10-100\un{MHz} regime will additionally permit
observation of the geomagnetic radio emission from the air shower. 
Yielding complementary information on the shower development a
triple-technology array consisting of radio sensors, the ground sampling
stations of IceTop and the in-ice optical
modules of IceCube, should significantly improve the understanding of cosmic
rays, as well as enhance many aspects of the physics reach of the observatory.
Here we present first results from two exploratory setups deployed at the South
Pole. Noise measurements from data taken in two consecutive seasons show a very
good agreement of the predicted and observed response of the antennas designed
specifically for this purpose.  The radio background is found to be highly
dominated by galactic noise with a striking absence of anthropogenic radio
emitters in the frequency band from 25-300\un{MHz}. Motivated by the excellent
suitability of the location, we present first performance
studies of a proposed Radio Air-Shower Test Array (RASTA) using detailed
MonteCarlo simulation and discuss the prospects for its installation.
\end{abstract}

\maketitle


\section{Introduction}
Detection of air-showers through geomagnetic radio emission in the \un{MHz}
regime offers various advantages compared to other detection technologies. In
contrast to ground-based shower sampling, the radio signal represents the integral
shower development and hence promises excellent energy resolution. In contrast
to fluorescence measurements, high duty-cycles can be achieved with a low-cost
technology. Further, the lateral distribution of the radio signal will
provide an independent measure of the depth of the shower maximum. Consequently,
radio extensions are now being developed in many existing air-shower
experiments.

The IceCube observatory at the South Pole employs a 
1\un{km^3} volume, instrumented with an array of photo-sensors as an ice-Cherenkov
detector mainly targeted at the detection of astrophysical neutrinos. Installed
at a depth between 1450-2450\un{m}, it is also sensitive to muons from air-showers
above an energy at the surface of $\sim300\un{GeV}$. This detector is
complemented by a $1\un{km^2}$ surface array of ice-Cherenkov tanks that
sample electrons and muons with a threshold of $10\un{MeV}$ from air-showers.
Coincident detection of high-energy muons in ice and the shower particles on
the surface provides a unique mechanism to study cosmic ray composition from
the knee to the ankle with a primary threshold of $\sim300\un{TeV}$, albeit with 
a limited aperture of $0.3\un{km^2 sr}$. 

Extending this observatory with a large array of radio antennas can
significantly enhance the physics potential of this
observatory~\cite{Boser:2010sw}: adding a third detection technology with
complementary shower observables will permit significant reduction of the
systematic uncertainties in the composition measurement. Extending to several
\un{km^2}, it may serve as a veto for cosmic ray air-showers, and thus enhance the neutrino
sensitivity in the southern hemisphere. Inverting the veto logic, it may also
serve as an instrument for ultra-high energy gamma-rays by searching for
muon-poor air-showers. 

In this paper, we show results from exploratory studies of the South Pole for
the radio detection technique, as well as a first simulation of the performance
that can be achieved with a dedicated test array. We also discuss the
prospects for this approach.
\section{Site exploration}
To establish the feasibility of radio air-shower detection at the South Pole
beyond the proof of principle provided by the LOPES\cite{Falcke:2005tc},
Codalema\cite{Ravel:2012zz} and AERA\cite{Fliescher:2012zz} experiments, two
question applying to this particular location must be addressed: is it possible
to build antennas that operate reliably at the environmental low temperatures; and does the
ambient noise permit a sufficiently low trigger threshold.
\subsection{Antenna design\label{sec:antenna}}
\begin{figure}
  \includegraphics[width=\columnwidth]{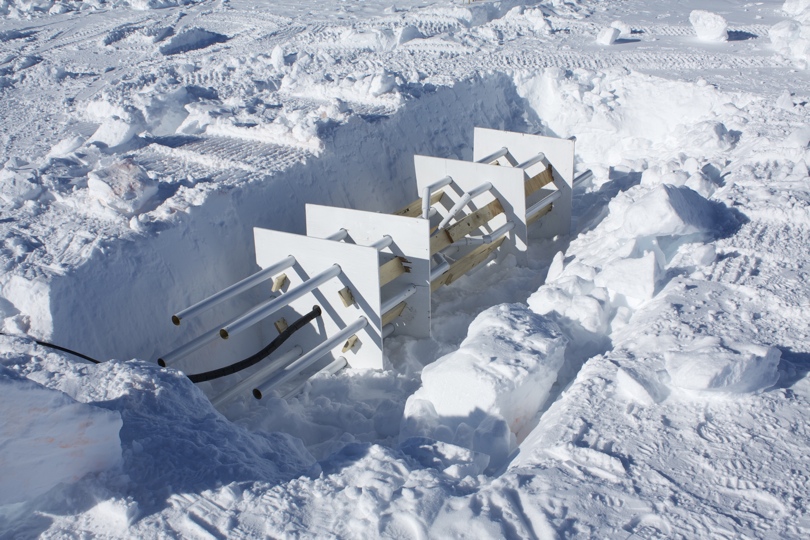}
  \caption{A fat tube-dipole deployed at the South Pole. Trenches are backfilled
  after deployment.\label{fig:deathray}}
\end{figure}

\begin{figure}[t!]
  \includegraphics[width=\columnwidth]{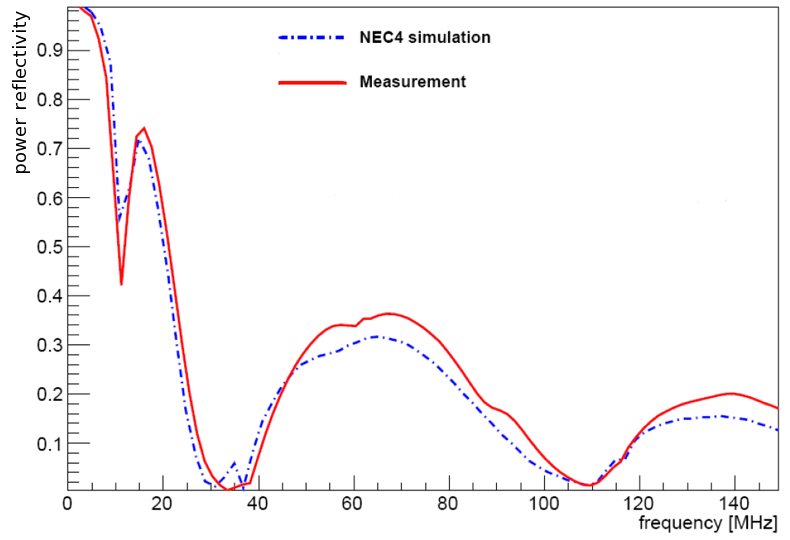}
  \caption{Power reflectivity as a function of frequency for a fat tube-dipole
  from lab measurements and simulation.\label{fig:powerrefl}}
\end{figure}

The harsh environment at the South Pole, in particular the low temperatures, pose
significant challenges to antenna construction. On the other hand, the good
radio transparency of the upper most layer of the glacier, or firn, allows to deploy antennas
where temperatures are stable at $-52 \pm 2\un{^\circ C}$ throughout the year
and they are not affected by occasional high winds. A number of designs have
been studied for this environment \cite{Vehring:2012} using NEC4 simulations. It
was found that the \emph{fat tube-dipole}, shown in Figure \ref{fig:deathray},
offers a low power reflectivity in the frequency range above 25\un{MHz}
as well as a low group delay, matching the predictions from NEC4
simulations well (c.f. Figure \ref{fig:powerrefl}). Manufactured from 5\un{cm} copper
tubes with a shrinking diameter towards the feedpoint, this design has proven
very robust for the six antennas of this type deployed at the
South Pole thus far.
\subsection{Experimental setups}
To study the radio environment for air-shower detection, two
experimental setups have been installed in the 2010/2011 and
2011/2012 polar seasons. Co-deployment with the ARA installations, and use of
their DAQ system, allowed for a minimum
investment. In the first season, two fat tube-dipole antennas have been deployed
at a distance of 30.5\un{m} and at a relative azimuthal orientation of
21\un{^\circ}. Both antennas were buried in trenches at a depth of $\sim
0.5\un{m}$, and the trenches were backfilled with snow. The antennas are connected to
the ARA-testbed DAQ system~\cite{Ara-Collaboration:2012fk}, where they are read out with a sampling rate
of up to 2\un{GS/s} at 12 bit and a bandwidth of 30-300\un{MHz}. In this setup, no
dedicated trigger is available for these two surface antennas, so only
events from a {\em forced trigger} every 2\un{s} enter the analysis. In the second
setup, four fat tube-dipole antennas were deployed, approximately in an equilateral
triangle with one antenna in the center. Three of these antennas are aligned
east-west, within $\pm3\un{^\circ}$ of the magnetic field, while one antenna
is offset to that direction by +22\un{^\circ}. To avoid different signal
propagation times, all antennas are connected to the central DAQ by 200\un{m}
of highly shielded cables and an overall gain of 74\un{dB}. In this setup
the signal from the four surface antennas in the 25\un{MHz}-120\un{MHz}
band is diplexed with the signal above 120\un{MHz} from the V-pol antennas
deployed in the ice for readout. In addition to the in-ice and minimum bias triggers,
a dedicated {\em surface trigger} has been operated in a 3-out-of-4 mode with individual
thresholds adjusted per channel. Due to low-temperature
instability of the commercial single-board computer employed in the DAQ system,
only a limited amount of data has been collected with this setup
from Jan, 26\textsuperscript{th} to May, 3\textsuperscript{rd} 2012, with an uptime of less than 50\%.
\subsection{Noise measurement}
\begin{figure}
  \includegraphics[width=\columnwidth]{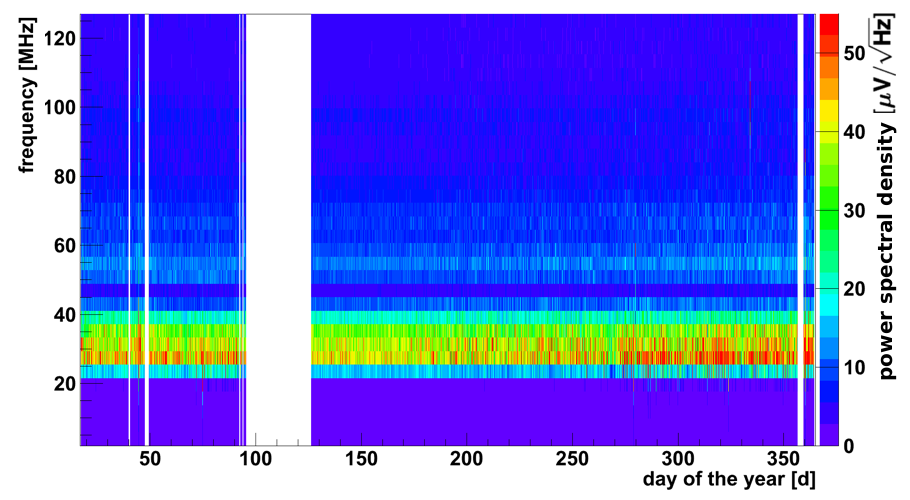}
  \caption{Signal power spectrogram as a function of days in 2011 for minimum
  bias events taken with the ARA-testbed setup.\label{fig:minbiasnoise}}
\end{figure}

\begin{figure}[t!]
  \includegraphics[width=\columnwidth]{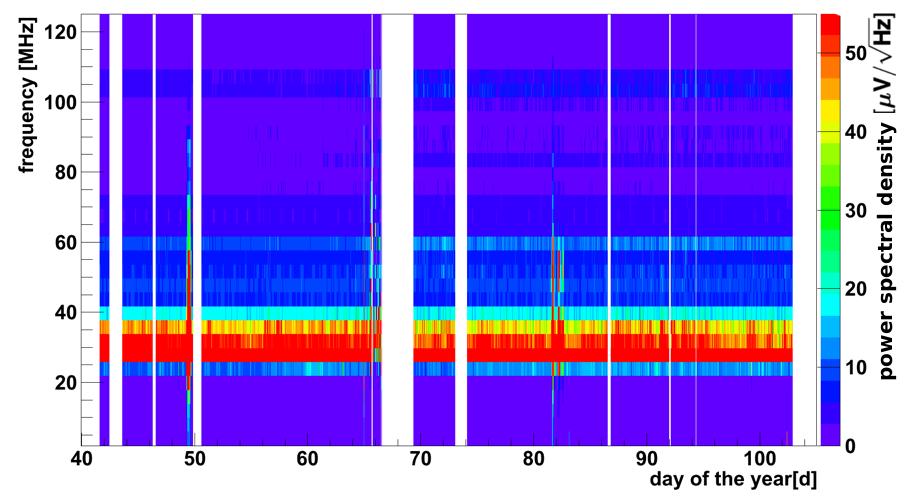}
  \caption{Signal power spectrogram as a function of days in 2012 for surface
  triggered events taken with the ARA-01 setup.\label{fig:trignoise}}
\end{figure}

Figure \ref{fig:minbiasnoise} shows the power spectral density, averaged over all
waveforms, obtained in one hour as a function of the day from the forced trigger
data in 2011. The signal is dominated by the falling spectrum of galactic noise
from the low-pass boundary at 25\un{MHz} up to about 90\un{MHz}, above which it is limited by thermal noise in the DAQ chain. A
single notch filter at 46.4\un{MHz} is used to mitigate the signal of the meteor
radar~\cite{meteorradar}. The absence of any other features and the very low
overall levels reflect the stability of the conditions at South Pole and the
absence of any steady-mode radio transmitters. Figure \ref{fig:trignoise}
shows the power spectral density averaged over one hour for the {\em surface
triggered} data obtained in 2012. The enhancement of the galactic noise band and
the absence of contributions in the frequency range above 50\un{MHz} indicates
that most triggers are caused by fluctuations in the thermal noise. Broadband
emission not consistent with galactic noise is only observed for three
short time ranges, possibly associated with exceptional air-craft communication
activity. This absence of transient broadband emitters makes South Pole a unique
location for the radio detection of air-showers. 

We have also studied the correlation of the noise with wind speed, air pressure
on the ground and air
temperature and found no significant dependence. A dependence
of the trigger rate on the wind speed, as previously observed with radio antennas
deployed with the RICE setup \cite{Kravchenko:2011im}, has not been confirmed.
This supports the hypothesis that these arise from discharge events on
structures above the firn surface that are absent in the setup discussed here.

\begin{figure}[h!]
  \includegraphics[width=\columnwidth]{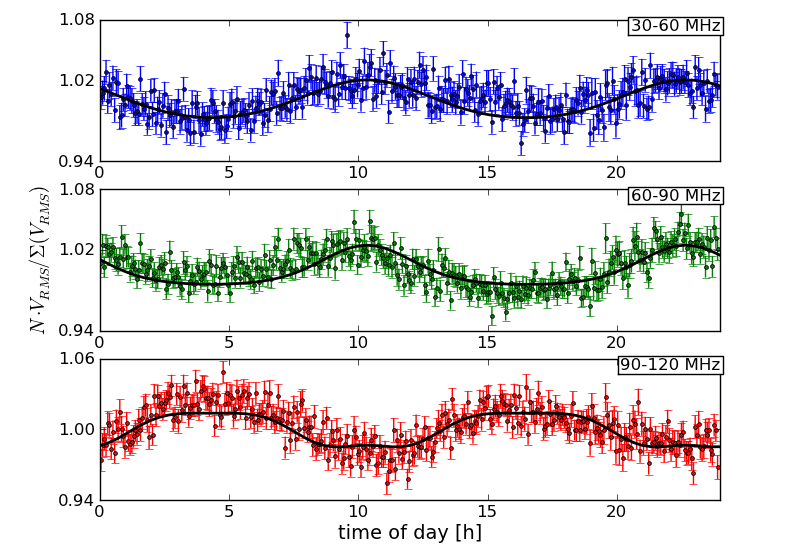}
  \caption{Predicted (lines) and measured (dots) normalized noise
  RMS amplitude variation for three different frequency ranges as a
  function of the time of the sidereal day.\label{fig:noisebands}}
\end{figure}

To establish the overall calibration of the system we have used the galactic
noise model by de~Olivera~\cite{deOliveiraCosta:2008pb}, that provides
interpolated full-sky intensity maps in the range from 10\un{MHz} to 10\un{GHz}.
Convolving these maps for each frequency with the antenna response taken from the
NEC4 simulation described in section \ref{sec:antenna}, we obtain a noise
spectrum prediction. Noise values for each component in the DAQ system are taken
from manufacturer information and are traced through the full chain. Above
25\un{MHz}, the resultant spectrum agrees to within $\pm2\un{dB}$ with the measured
spectrum. Since the antenna sensitivity is not isotropic in azimuth, this
agreements only holds if averaging the noise spectra over time periods much
longer than one day. By convolving the galactic noise prediction with the
antenna response instead for a particular antenna orientation, {\it i.e.} a particular time of
the day, a time dependant noise level is expected in each frequency band. Figure
\ref{fig:noisebands} shows the variation in noise RMS amplitudes for three
frequency bands normalized to the daily average in each band. Higher
noise levels are predicted in the lower frequency band when the antenna is
aligned perpendicular to the direction of the galactic center (for
co-polarized radio-waves). However, for the highest frequency band, the largest
noise levels are predicted when the antenna is oriented towards the galactic
center (in cross-polarized mode). The inversion point, defined as the frequency
at which the antenna becomes more sensitive in cross-polarized than in
co-polarized mode, as well as the amplitude of the azimuthal modulation of the
antenna response, function will depend on the electrical permittivity of the surrounding
snow. From a $\chi^2$ fit procedure we find the best agreement with the measured
noise level variations for a permittivity of $\varepsilon = 1.4$ at
$\chi^2/\rm{NDF}=1.31$. Different approximations have been proposed for the
dependence of the permittivity of snow on the snow density~\cite{dielect:1980,
dielect:1985}, suggesting a corresponding snow density of $\rho_{\rm snow} =
0.21-0.29\un{g/cm^3}$. Values of $\rho_{\rm snow} = 0.4\un{g/cm^3}$ have been
measured for pristine snow at the South Pole at a depth of one meter, but lower values
are plausible for the backfilled trenches in which the antennas are located.
\section{Performance estimate}
\begin{figure}[b!]
  \includegraphics[width=\columnwidth]{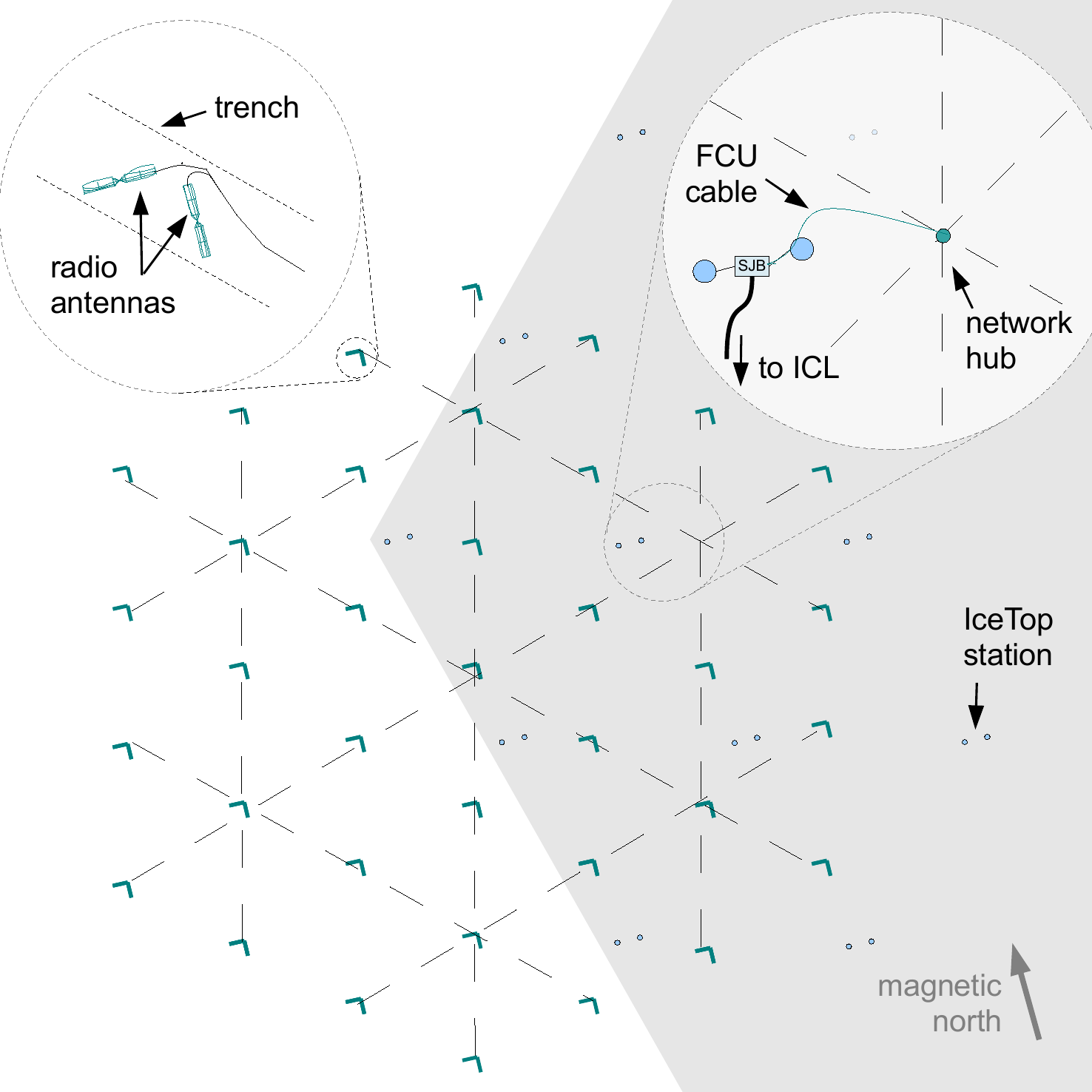}
  \caption{Proposed layout of the Radio Air-Shower Test Array (RASTA)\label{fig:rastageo}}
\end{figure}
Given the ideal environmental conditions outlined above, a Radio
Air-Shower Test Array (RASTA) has been proposed to establish
the feasibility of radio air-shower detection and study the physics potential~\cite{Boser:2010sw}. A dedicated simulation has
been developed to study the response. The proposed detector geometry, shown in figure
\ref{fig:rastageo}, consists of 74 antennas arranged in pairs in a hexagonal grid
with a spacing of 62.5\un{m}, half the grid size of the IceTop array.
In each pair, one antenna is oriented north-south and
the other east-west relative to the geomagnetic field. 
\subsection{Simulation chain}
\begin{figure}
  \includegraphics[width=\columnwidth]{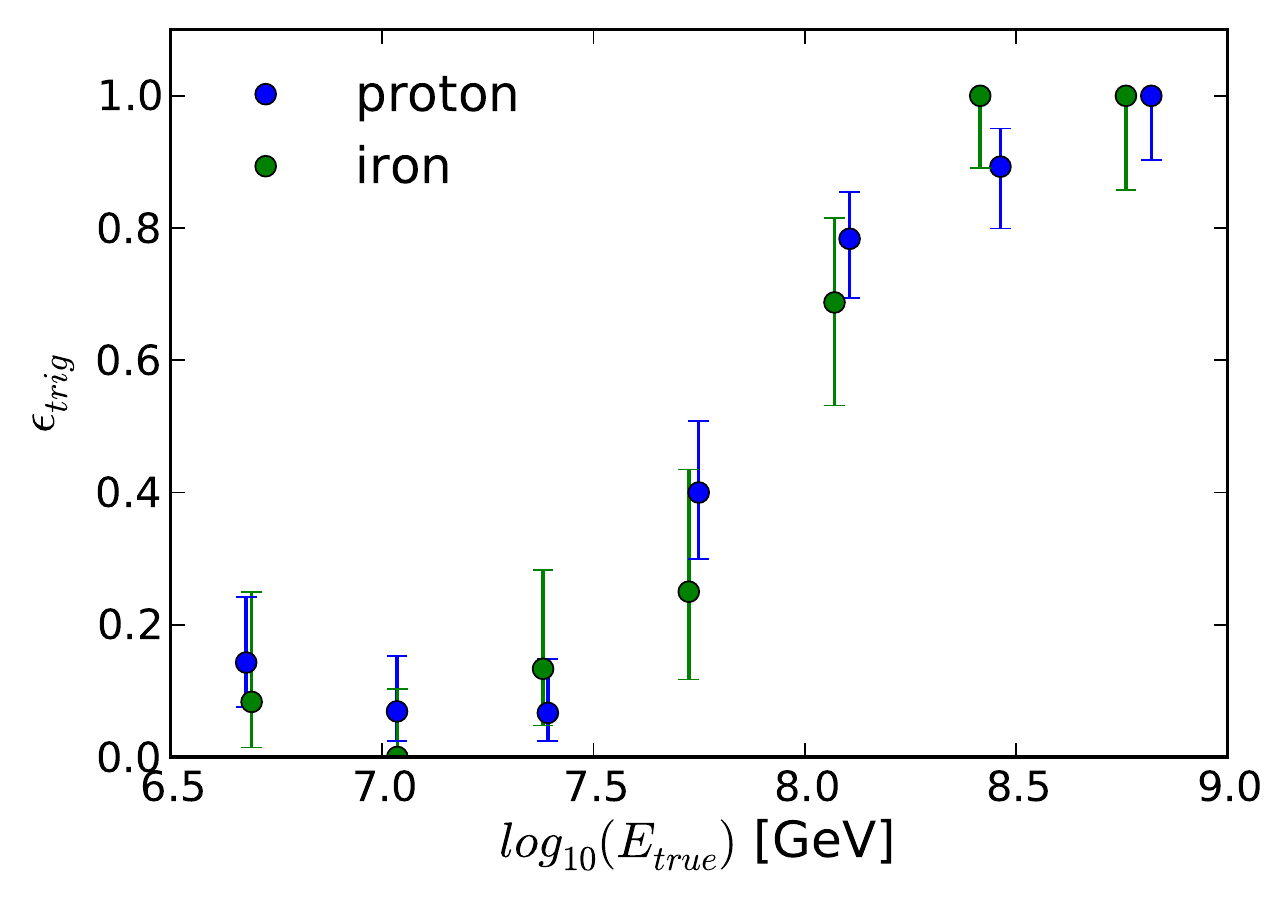}
  \caption{Trigger efficiency as a function of energy for proton (green) and
  iron (blue) primaries.\label{fig:threshold}}
\end{figure}

Three different emission models have been used as inputs for the simulation
chain: MGMR~\cite{Scholten:2010na}, REAS3.0~\cite{Ludwig:2010pf} and
CoREAS~\cite{coreas:2012}. For each model, we have simulated several hundred events in
the energy range between 3\un{PeV} to 1\un{EeV} and with zenith angles up to
60\un{^\circ}. The electric field as a function of time is sampled at each
antenna pair location and folded with the antenna response (c.f. section
\ref{sec:antenna}). Galactic noise, following the model of
Cane~\cite{cane:1979}, and
thermal noise are added assuming a feedpoint impedance of 50\un{\Omega}. We
simulate a second order bandpass filter from 25-300\un{MHz} and assume
digitization with 300\un{MHz} at 16-bit resolution. An event is defined to
trigger the array if 4 antennas in 4 distinct pairs have a signal 5 times above
the thermal noise within a time window of 1\un{\mu{}s}. A sample
trace of the simulated signal after digitization is shown in Figure
\ref{fig:hilbert}. Figure \ref{fig:threshold} shows the trigger efficiency for
proton and iron primaries
from the CoREAS simulation which includes refractive index variation throughout the
atmosphere. A 50\% efficiency is reached at $\sim100\un{PeV}$, while for the
REAS3.0 simulation, that assumes a constant refractive index, we find a 50\%
efficiency at $\sim30\un{PeV}$. It should be noted that, due to the strong
dependence of the signal strength on the orientation w.r.t to the Earth's magnetic
field and the zenith angle, a fraction of the events will also be detected at
energies well below this threshold and with a steeply falling cosmic ray
spectrum that may dominate the accumulated data.
\subsection{Event reconstruction}
\begin{figure}[t!]
  \includegraphics[width=.9\columnwidth]{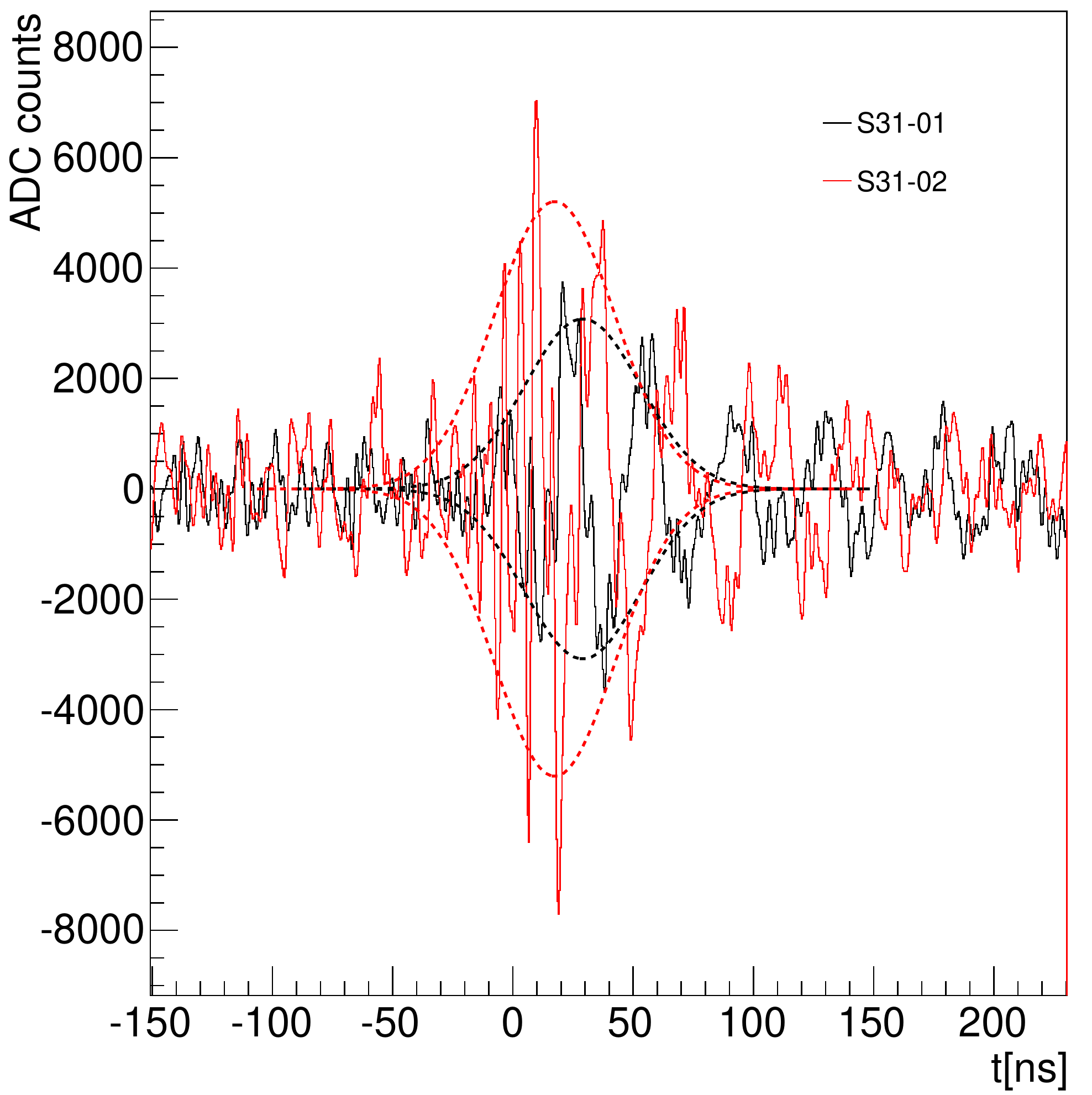}
  \caption{Simulated signal traces in one antenna pair from REAS3.0 including noise and system
  response. Time and amplitude are extracted from a Gaussian fit (dashed line)
  to the Hilbert envelope.\label{fig:hilbert}}
\end{figure}

\begin{figure}[t!]
  \includegraphics[width=.9\columnwidth]{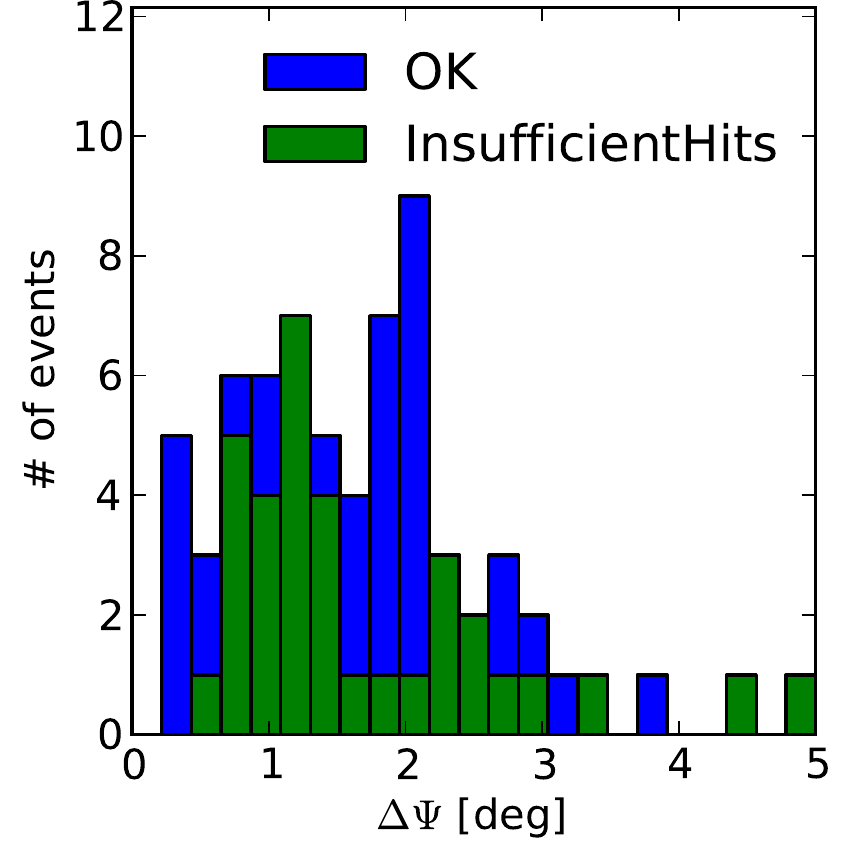}
  \caption{Angular resolution from a plane wave fit that iteratively rejects hits
  with large pull factors. Iteration will stop if the fit converges (blue) or
  not enough hits remain (green).\label{fig:angres}}
\end{figure}

In a first attempt of event reconstruction, we employ an iterative plane wave
fit procedure. We perform a Hilbert transform to the signal in each
antenna and fit a Gaussian to the peak (c.f.~Figure~\ref{fig:hilbert}). The peak time and amplitude of this
Gaussian fit are used as a proxy to define a signal \emph{hit} if the
amplitude exceeds five times the RMS amplitude in the trace. Using singular
value decomposition (which is faster and more accurate than $\chi^2$
minimization), we
obtain the best plain wave best match to the set of antenna positions and arrival
times. Excluding one of the single antenna hits at a time, we find the hit with
the largest pull on the reconstructed direction and recursively remove these
hits until the resulting direction does not vary by more than 0.3\un{^\circ} (or
not enough hits are left to continue the procedure). Figure \ref{fig:angres}
shows the resulting angular resolution for CoREAS simulated events. A median
angular resolution of $1.5\un{^\circ}$ can be achieved even with this simple
approach, comparable to the resolution of IceTop. We also use the amplitude
weighted geometric mean of all hits as a proxy for the shower core location, and
find an average resolution of $\sim10\un{m}$ for REAS3.0 simulations not
including refractive index variations. This simple proxy does not work for
CoREAS simulation. With refractive index effects included, the lateral
distribution no longer falls monotonically with the radius for small
distances to the shower axes~\cite{Werner:2012cr}.
\section{Summary and Outlook}
A series of experimental setups has been installed at the South Pole to explore
the suitability of the location for an air-shower detector based on radio
technology. Noise measurements from more than a year of data-taking demonstrated the
excellent suitability of the South Pole for this approach. A detailed comparison of
the measured noise spectra show agreement with model predictions on the few
\un{dB} level, and hence a profound understanding of the antenna response and DAQ
chain calibration. Motivated by this success, proposals have been submitted to
the BMBF and the NSF for an extended radio air-shower test array --- RASTA --- consisting of 74
antennas deployed over a few 100\un{m^2} area. Performance estimates using
particle-level shower simulations, as well as detailed simulation of the galactic
noise, the DAQ response and antenna response show that a 50\% detection
efficiency at $\sim 100\un{PeV}$ may be achievable in such a geometry for a
simple threshold-based trigger. Using straightforward plane-wave reconstruction
methods, with iterative signal cleaning algorithms, suggest that angular
resolutions of about 1$\un{^\circ}$ and shower core reconstructions of 10\un{m}
may be obtained. This performance is comparable to the performance that has been measured for the
IceTop detector. While successfully funded by the BMBF in 2011, funding for
logistic support for such an installation at the South Pole has remained
elusive, temporarily delaying future research for this promising project.




\bibliographystyle{aipproc}   

\bibliography{rasta}

\IfFileExists{\jobname.bbl}{}
 {\typeout{}
  \typeout{******************************************}
  \typeout{** Please run "bibtex \jobname" to optain}
  \typeout{** the bibliography and then re-run LaTeX}
  \typeout{** twice to fix the references!}
  \typeout{******************************************}
  \typeout{}
 }

\end{document}